\documentclass[12pt]{article}

\begin{document}

\title{Nonlinear Optical Vector Amplitude Equations. Polarization and
Vortex Solutions}

\author{Lubomir M. Kovachev\\
Institute of Electronics, Bulgarian Academy of Sciences,\\
Tzarigradcko shossee 72,1784 Sofia, Bulgaria,
\and
David R. Andersen\\
Department of Electrical and Computer Engineering,\\
 Department of Physics and Astronomy, \\
 The University of Iowa, Iowa City, IA 52242, USA}
\maketitle
\begin{abstract}
We investigate two kind of polarization of localized optical waves in
nonlinear Kerr type media, linear and combination of linear and circular.
In the first case of linear polarized components we have obtained the vector
version of 3D+1 Nonlinear Schr\"{o}dinger Equation (VNSE).
We show that these equations admit exact vortex solutions with spin $\ell =1$.
We have determined the dispersion region and medium parameters necessary for
experimental observation of these vortices in the conlusion. In the second
case we represent the electric and magnetic fields as a sum of circular and
linear components. We supouse also that our nonlinear media in this case admit
linear magnetic polarization. This allows us to reduce the Maxwell's
equations to a set of amplitude Nonlinear Dirac Equations (NDE).  We find two
representations on NDE- spherical and spinor. In the spherical representation
we obtain optical votices with spin $\ell =1$ and in the spinor
representation, vortices with spin $\ j=1/2.$
\end{abstract}

\section{Introduction}

The scalar theory of optical vortices was based on the well-known 2D+1
paraxial Nonlinear Schr\"{o}dinger equation (NSE) \cite{SW,CH}. A
generalization of the scalar paraxial theory of optical vortices based on the
investigation of co-called spatio-temporal evolution equations
\cite{SIL,ED,MLD} has also been performed. The polarization and the vector
character of the electric field play an important role in dynamics and the
stabilization of the vortices. In \cite{LW} this problem was
discussed for the first time and the existence of polarized vortices was
predicted. To investigate this phenomenon more completely, it is necessary to
investigated the corresponding vector amplitude equations. In this paper, we
investigated two type of vector amplitude equations, VNSE for three linear
othogonal polarized component, and a Dirac representation of Nonlinear
Maxwell's equations for a combination of linear and circular polarized
components.

\section{ Vector NSE}

The vector version of 3D+1 amplitude equation of electrical field,
describing the propagation of light in dispersion media with Kerr
nonlinearity in a dimensionlles coordinate system, moving with group
velocity \ my be writen:

\begin{equation}\label{eq1}
2i\alpha\frac{\partial \vec{A}}{\partial t} +\Delta_{\perp}\vec{A}
-\beta {\frac{{\partial^{2}\vec{A}}}{{\partial z^{2}}}}+\gamma {{\left|{
\vec{A}}\right| }^{2}}\vec{A}=0,
\end{equation}
where $\vec A$ is the normed amplitude of the electrical field,
$\alpha=kr_{0}^{2}/t_{0}v;$ $\beta =v_{i}^{2}k^{"}k;$
$\gamma =k^{2}r_{0}^{2}n_{2}{\left| {A}\right| }_{0}^{2};$
are corresponding normed constant, $v$ is
group velocity, $k$ is carrying wave number, $r_{0}$ and $t_{0}$ are the
spatial and time dimensions of amplitude field, $k"$ is the dispersion
parameter, $n_{2}$ is the nonlinear refractife index. We investigate the
case $\beta =-1;\gamma =1$. The constant $\alpha $ has a tipical value of
$\alpha \approx{10^{2}-10^{3}} $ in the optical region ($\alpha
\approx{r_{0}k}$).

To solve the vector equation (\ref{eq1}), we use the method of
separation of variables. The amplitudes of three linear polarized orthogonal
vector fields $\vec{A}^{j}(x,y,z,t)$ \ are represented as:

\begin{equation} \label{eq2}
\vec{A}(x,y,z,t)=\sum_{\vec{j}=\vec{x},\vec{y},\vec{z}}
{\vec{j} A^{j}(x,y,z)\exp (-i{\alpha }t)}.
\end{equation}
In a Cartesian coordinate system, for solutions of the form of (\ref{eq2}),
the vector equation (\ref{eq1}) is reduced to a scalar system of three
nonlinear wave equations:

\begin{equation}\label{eq3}
\alpha ^{2}A^{l}+\Delta A^{l}+{\sum\limits_{j=x,y,z}{\left( {{\left| {A^{j}}
\right| }^{2}}\right) }}A^{l}=0;\quad l=x,y,z.
\end{equation}
The system (\ref{eq3}) can be written in spherical variables:

\begin{equation} \label{eq4}
\alpha ^{2}A^{l}+\Delta _{r}A^{l}+{\frac{{1}}{{r^{2}}}}\Delta _{\theta
,\varphi }A^{l}+{\sum\limits_{j=x,y,z}{\left( {{\left| {A^{j}}\right| }^{2}}
\right) }}A^{l}=0,
\end{equation}
where $l=x,y,z$. Next we represent the components of the field as a product of a radial and
an angular part:

\begin{equation}\label{eq5}
A^{i}(r,\theta ,\varphi )=R(r)Y_{i}(\theta ,\varphi );i=x,y,z,
\end{equation}
with the additional constraint on the angular parts:

\begin{equation}\label{eq6}
\left| Y_{x}(\theta ,\varphi )\right| ^{2}+\left| Y_{y}(\theta ,\varphi
)\right| ^{2}+\left| Y_{z}(\theta ,\varphi )\right| ^{2}=const..
\end{equation}
Multiplying each of the equations (\ref{eq4}) by the
$\frac{r^{2}}{RY_{i}}$, and bearing in mind the constraint expressed
in (\ref{eq6}), we obtain:

\begin{equation}\label{eq7}
{\frac{{r^{2}\Delta _{r}R}}{{R}}}+r^{2}\left( {\alpha ^{2}+{{\left| {R}
\right| }^{2}}}\right) =-{\frac{{\Delta _{\theta \varphi }Y_{j}}}{{Y_{j}}}}
=\ell (\ell +1),
\end{equation}
where $\ell $ is one number. Thus the following equations for the radial and
the angular parts of the wave functions are obtained:

\begin{equation} \label{eq8}
{\Delta _{r}R}+{\alpha ^{2}R+{{\left| {R}\right| }^{2}}}R-\frac{\ell (\ell
+1)}{r^{2}}=0
\end{equation}

\begin{equation} \label{eq9}
{\Delta _{\theta \varphi }Y_{j}+\ell (\ell +1)=0;j=x,y,z}.
\end{equation}
As equations (\ref{eq8}),(\ref{eq9}) shows, the nonlinear
term occurs only in the radial components of the fields,while for the angular
part we have the usual linear eigenvalue problem. Going back to set
(\ref{eq4}), we see that the separation of variables for the spherical
functions, which satisfied condition (\ref{eq6}), is posible only for
$\ell=1$.

\begin{equation}\label{eq10}
{Y_{x}} ={Y}_{1}^{-1}=\sin \theta \cos \varphi ;
{Y_{y}} ={Y}_{1}^{1}=\sin \theta \sin \varphi ;
{Y_{x}} ={Y}_{1}^{0}=\cos \theta.
\end{equation}
By choosing one of these angular components for each of the field components
we see that the eigenfunctions (\ref{eq10}) are solutions to the angular
part of the set of equations (\ref{eq4}). The nonlinear radial part of
equations admit exact ''de Brogile soliton'' solutions \cite{BR}

\begin{equation}\label{eq11}
R=\frac{\sqrt{2}}{i}\frac{e^{i\alpha r}}{r}.
\end{equation}

The real solutions of the vector amplitude equation (\ref{eq4}) in a
fixed basis is then:

\begin{displaymath}
A_{x}=\Re\left(R(r)Y_{x}\right) =\sqrt{2}\frac{\sin (\alpha r)}{r}
\sin \theta \cos\varphi \cos (\alpha t)
\end{displaymath}

\begin {equation}
A_{y}=\Re\left(R(r)Y_{y}\right)=\sqrt{2}\frac{\sin (\alpha r)}{r}
\sin \theta \sin\varphi \cos (\alpha t)
\end {equation}

\begin{displaymath}
A_{z}=\Re\left(R(r)Y_{z}\right)=\sqrt{2}\frac{\sin (\alpha r)}{r}
\cos \theta \cos (\alpha t).
\end{displaymath}

\section{ Nonlinear Maxwell's Equations}

We consider the Maxwell's equations for the case a source-free nonlinear
medium with linear and nonlinear electric polarization and linear
magnetic polarization (magnetization) \cite{AH}, \cite{LV}:

\begin{equation}\label{eq12}
\nabla\times{\vec E}=-\frac{1}{c}\frac{\partial\vec B}{\partial t};
\nabla\times{\vec B}=\frac{1}{c}\frac{\partial\vec D}{\partial t}
\end{equation}

\begin{equation}\label{eq13}
\nabla\cdot\vec D=0;\nabla\cdot\vec B=\nabla\cdot\vec H=0
\end{equation}

\begin{equation}\label{eq14}
\vec D=\vec P_l+4\pi\vec P_{nl}
\end{equation}

\begin{equation} \label{eq15}
\vec B=\vec H+4\pi\vec M_l,
\end{equation}
where $\vec E$ and $\vec H$ are the electric and magnetic intensity fields,
$\vec D$ and $\vec B$ are the electric and magnetic induction fields,
$\vec P_l$, $\vec P_{nl}$ are the linear and nonlinear polarization of the
medium respectively and $\vec M_l$ is the linear magnetic polarization.
We apply the slowly varying amplitude approximation to the Maxwell equations
(\ref{eq12})-(\ref{eq15}):

\begin{equation} \label{eq17}
\vec{E}(x,y,z,t)=\vec{A}(x,y,z,t)e^{i\omega _{0}t}
\end{equation}

\begin{equation} \label{eq18}
\vec{H}(x,y,z,t)=\vec{C}(x,y,z,t)e^{-i\omega _{0}t},
\end{equation}
where $\vec{A},$ $\vec{C}$ and $\omega _{0}$ are the amplitudes of the
electric and magnetic fields and the optical frequency respectively.
After using the Fourier representation and condition of slowly-varying
amplitudes we obtained the next system of Nonlinear Maxwell amplitude
Equations(NME), writed in rescaled variables:

\begin{equation} \label{eq19}
\nabla \times \vec{A}=i\alpha_2 \vec{C}-\delta\frac{ \partial \vec{C}}
{\partial t}-\beta_2\frac{\partial ^{2} \vec{C}}{\partial t^{2}}
\end{equation}

\begin{equation} \label{eq20}
\nabla \times \vec{C}=i\alpha_1 \vec{A}+\frac{\partial \vec{A}}{\partial t}
-i\beta_1 \frac{\partial ^{2}\vec{A}}{\partial t^{2}}+i\gamma
|\vec{A}|^{2}\vec{A} \end{equation}

\begin{equation} \label{eq21}
\nabla \cdot \vec{A}=0 ;\nabla \cdot \vec{C}=0,
\end{equation}
where the constants are
$k_{1}=\frac{w\epsilon(\omega)}c$; 
$k_{2}=\frac{w\mu(\omega)}c$;
$\frac 1v_{i}=\frac{\partial k_i}{\partial\omega}$;
$\alpha_i =k^0_{i}r_0$;
$\beta_i =k^{"}_{i}r_{0}/2t_{0}^{2}$;
$\gamma=r_{0}k_{1}n_{2}\left| A_{0}\right| ^{2}$;
$\delta=\frac {v_1}{v_2}$.
While the nonlinear condition $\gamma =1$ can be satisfied, the constants
$\alpha_i $; $i=1,2$ have typical values
$\alpha_i \approx 10^{2}(\alpha_i \approx r_{0}k_{i})$. The
constants $\beta_i $ have typical values
$\beta_1\approx 10^{-5}-10^{-6}<<1$;$\beta_2<\beta_1$ and for picosecond
and sub-picosecond pulses in the transparency region of nonlinear optical
media they may be neglected.

\section{ Dirac representation of NME}

We will neglecting the second derivative in time terms of the
eqn.(\ref{eq19})-(\ref{eq20}), as $\beta_i<<1$. We solve the NME
(\ref{eq19})-(\ref{eq21}), by the method of separation of variables.
The slowly varying amplitude vectors of the electric field $\vec{A}$ and
the magnetic field $\vec{C}$ are represented as:

\begin{equation} \label{eq26}
\vec{A}(x,y,z,t)=\vec{F}(x,y,z,t)e^{i\triangle \alpha t}
\end{equation}

\begin{equation} \label{eq27}
\vec{C}(x,y,z,t)=\vec{G}(x,y,z,t)e^{i\triangle \alpha t}.
\end{equation}
Substituting these forms into (\ref{eq19})-(\ref{eq21}), we obtain the
stationary version of the NME:

\begin{equation} \label{eq28}
\nabla \times \vec{F}=-i\vartheta_2 \vec{G};
\nabla \times \vec{G}=i\vartheta_1 \vec{F}+i\gamma |\vec{F}|^{2}\vec{F}
\end{equation}

\begin{equation} \label{eq29}
\nabla \cdot \vec{F}=0 ;\nabla \cdot \vec{G}=0,
\end{equation}
where $\vartheta_1 =\alpha_1 +\triangle \alpha $;
$\vartheta_{2}=\delta\triangle\alpha -\alpha_2 >0$ . When the electric and
magnetic fields are represented as a sum of a linear and circular polarized
component it is possible to reduce eqns.(\ref{eq28})- (\ref{eq29}) to a
system of four nonlinear equations. Substituting \cite{DAC}:

\begin{equation} \label{eq31}
\Psi _{1}=iF_{z} ;\Psi _{2}=iF_{x}-F_{y};\Psi_{3}=-G_{z};
\Psi _{4}=-G_{x}-iG_{y}
\end{equation}
into the nonlinear system (\ref{eq28})-(\ref{eq29}) we obtain a stationary
nonlinear Dirac system (NDE):

\begin{equation} \label{eq32}
\left ( \frac{\partial }{\partial x}-i\frac{\partial }{\partial y}\right)
\Psi _{4}+\frac{\partial }{\partial z}\Psi _{3}=-i(\vartheta
_{1}+\gamma \sum\limits_{j=1}^{2}|\Psi _{j}|^{2}) \Psi_{1}
\end{equation}

\begin{equation} \label{eq33}
\left ( \frac{\partial }{\partial x}+i\frac{\partial }{\partial y}\right)
\Psi _{3}-\frac{\partial }{\partial z}\Psi _{4}=-i( \vartheta _{1}+\gamma
\sum\limits_{j=1}^{2}|\Psi _{j}|^{2}) \Psi_{2} \end{equation}

\begin{equation} \label{eq34}
\left( \frac{\partial }{\partial x}-i\frac{\partial }{\partial y}\right)
\Psi _{2}+\frac{\partial }{\partial z}\Psi _{1}=-i\vartheta _{2}\Psi _{3}
\end{equation}

\begin{equation} \label{eq35}
\left( \frac{\partial }{\partial x}-i\frac{\partial }{\partial y}\right)
\Psi _{1}-\frac{\partial }{\partial z}\Psi _{2}=-i\vartheta _{2}\Psi _{4}.
\end{equation}
As may be noted, the {\it optical} NDE are significantly
different from the NDE in field theory. The nonlinear part appears {\it only}
in the first two coupled equations of the system.

\section{Vortex solutions with spin $l=1$}

The NDE (\ref{eq32})-(\ref{eq35}) have both spherical and spinor
symmetry only in the case where the nonlinear part does not manifest an
angular dependence on the radial variable
$\sum\limits_{j=1}^{2}|\Psi_{j}|^{2}=F(r)$.
This type of solution may be found using with the following technique: using
Pauli matrices, we write the NDE system (\ref {eq32})-(\ref{eq35}) as:

\begin{equation} \label{eq36}
(\vec{\sigma}\cdot \vec{P}) \phi =-i( \vartheta _{1}+\gamma
\sum\limits_{j=1}^{2}|\eta _{j}|^{2})\eta
\end{equation}

\begin{equation} \label{eq37}
(\vec{\sigma}\cdot \vec{P}) \eta =-i\vartheta _{2}\phi,
\end{equation}
where
\begin{math}
\vec{\sigma}
\end{math}
are the Pauli matrices,
\begin {math} \vec{P}=\left(\frac{\partial }{
\partial x};\frac{\partial }{\partial y};\frac{\partial }{\partial z}\right)
\end {math}
are the differential operator and $\eta$ and $\phi$
are the corresponding spinors. After substituting eqn. (\ref{eq37}) into
eqn. (\ref{eq36}) we obtain:

\begin{equation} \label{eq38}
( \vec{\sigma}\cdot \vec{P}) ( \vec{\sigma}\cdot \vec{P}
) \eta =-\vartheta _{2}( \vartheta _{1}+\gamma
\sum\limits_{j=1}^{2}|\eta _{j}|^{2}) \eta
\end{equation}
When there is no external electric or magnetic field, the operator on the
left-hand side of eqn. (\ref{eq38}) is simply the Laplacian operator
$\triangle$. From (\ref{eq38}) and we obtain:

\begin{equation} \label{eq40}
\vartheta _{1}\vartheta _{2}\eta +\vartheta _{2}\gamma
\sum\limits_{j=1}^{2}|\eta _{j}|^{2}\eta +\triangle \eta =0
\end{equation}
The scalar variant of these equations has been investigated in many papers
but exact localized solutions have not been found. In the case $l=1$ we look
for spinors of the form:

\begin{equation} \label{eq43}
\eta_1=\widetilde{\eta}(r)\cos{\theta};
\eta_2=\widetilde{\eta}(r)\sin{\theta}\exp{i\phi}.
\end{equation}
After substituting eqn. (\ref{eq43}) into equations (\ref{eq40}) the
following equation describing the radial dependence is obtained:

\begin{equation} \label{eq44}
\vartheta _{1}\vartheta _{2}\widetilde{\eta }+\vartheta _{2}\gamma
\sum\limits_{j=1}^{2}\left| \widetilde{\eta }_{j}\right| ^{2}\widetilde{\eta
}+\frac{\partial ^{2}\widetilde{\eta }}{\partial r^{2}}+\frac{2}{r}\frac{
\partial \widetilde{\eta }}{\partial r}-\frac{2}{r^{2}}\widetilde{\eta }=0.
\end{equation}
The angular parts are the standard spherical harmonics with $l=1$. This
system has exact vortex {\it de Broglie} soliton \cite{BR} solutions of the
form:

\begin{equation} \label{eq45}
\widetilde{\eta }(r)=\frac{\sqrt{2}}{i}\frac{\exp (i\sqrt{\vartheta
_{1}\vartheta _{2}}r)}{r}
\end{equation}
if $\theta _{2}\gamma =1$.

\section{ Hamiltonian representation of the NDE: Orbital momentum, first
integrals and vortex solutions with spin $j=1/2$}

It is not difficult to show that for the NDE system of eqns.
(\ref {eq32})-(\ref{eq35})  a Hamiltonian of the form:

\begin{equation} \label{eq46}
H=\left( \vec{\sigma}\cdot \vec{P}\right) +\sum\limits_{j=1}^{2}\left| \Psi
_{j}\right| ^{2}
\end{equation}
may be written. Using this, NDE may be rewritten in the form:

\begin{equation} \label{eq47}
H\Psi =\varepsilon \Psi,
\end{equation}
where $\varepsilon =\left( -iv_{1},-iv_{1},-iv_{2},-iv_{2}\right) $ is the
energy operator. Here we investigate the case where the nonlinear part, as a
sum of square module of spinors, depends only on the radial component
$\sum\limits_{j=1}^{2}\left| \Psi _{j}\right| ^{2}=F(r)$.
We also introduce here the well known orbital momentum operator $\vec{L}$,
spin momentum $\vec{S}$, and the full momentum $\vec{J}$, as well as:

\begin{equation} \label{eq49}
\vec{L}=\vec{r}\times \vec{P};
\vec{J}=\vec{L}+\frac{1}{2}\vec{S}.
\end{equation}
It straightforward to show that the Hamiltonian of eqn. (\ref{eq46})
commutes with the operators $\vec{J}^{2}$ and $J_{z}$ (the z- projections
must be x or y). Using these symmetries and the condition that the
nonlinearity be centrosymmetric, we can solve the NDE equations (\ref{eq47})
by a separation of variables technique. To find vortex solution of the NDE (
\ref{eq47}) with spin $j=1/2$ we use the spinor representation of these
equations. We look for solutions of the form:

\begin{equation}
\Psi _{1} =a(r)\Omega _{jlm};\,\Psi _{2} =a(r)\Omega _{jlm}
\end{equation}
\begin{equation}
\Psi _{3} =ib(r)\Omega _{jl^{^{\prime }}m};\Psi _{4}=ib(r)\Omega
_{jl^{^{\prime }}m}
\end{equation}
where $\Omega _{jlm}$ is the spherical spinor, $l+l^{^{\prime }}=1$, and
$a(r)$ and $b(r)$ are arbitrary radial functions. Using the symmetries of (
\ref{eq47}) and the fact, that the nonlinear parts depend only on $r$, we
separate variables and obtain the following system of equations for the
radial part:

\begin{equation} \label{eq52}
\frac{\partial a(r)}{\partial r}+\frac{1+\sigma}{r}a(r)=-\vartheta
_{2}b(r)
\end{equation}

\begin{equation} \label{eq53}
\frac{\partial b(r)}{\partial r}+\frac{1-\sigma}{r}b(r)=\vartheta
_{1}a(r)+\gamma \left| a(r)\right| ^{2}a(r),
\end{equation}
where $\sigma =l(l+1)-j(j+1)-1/4$. We find localized solutions of
these equations only for the angular component corresponding to $l=1$ and $\
j=1/2$. In this case the system (\ref{eq52})-(\ref{eq53}) becomes:

\begin{equation} \label{eq54}
\frac{\partial a(r)}{\partial r}+\frac{2}{r} a(r)=-\vartheta_{2}b(r)
\end{equation}

\begin{equation}\label{eq55}
\frac{\partial b(r)}{\partial r}=\vartheta_{1}a(r)+
\gamma \left| a(r)\right| ^{2}a(r).
\end{equation}
As was shown above, this system has exact radial solutions of the form:

\begin{equation} \label{eq56}
a(r)=\frac{\sqrt 2}{i}\frac{\exp{i\sqrt{\vartheta_{1}\vartheta_{2}}r}}{r}
\end{equation}

\begin{equation}
b(r)=-\frac{\sqrt 2}{i\vartheta_{2}}\left(
\frac{\exp{i\sqrt{\vartheta_{1}\vartheta_{2}}r}}{r^2}+
\frac{i\sqrt{\vartheta_{1}\vartheta_{2}}
\exp{i\sqrt{\vartheta_{1}\vartheta_{2}}r}}{r}\right).
\end{equation}
The vortex solutions of the spinor representation
have four components and this type of solutions is found only for the case
where the spin $j=1/2$.

\section{ Conclusion}

In this paper we investigated two cases of vector amplitude equations
of electromagnetic field in nonlinear Kerr media. In the first case, for
three orthogonal linear polarized component of electrical field we obtained
exact vortex solutions of VNSE with spin $l=1$. The numerical analysis
shows, that these solutions are stable at distances comparable to those
where localized waves with the same amplitudes of the scalar 3D+1 NSE
self-focuced rapidly. The conditions $\beta=-1$ is satisfied only
near to Langmuir frequency in plasma or near to some of the electron
resonances. In the second case we have derived a set of Nonlinear Maxwell
amplitude Equations (NME) for nonlinear optical media with dispersion of the
magnetic susceptibility. We show that for the case of linear and circularly
polarised components of the electric and magnetic fields, the NME is reduced
to the Nonlinear Dirac system of equations (NDE).  Using the method of
separation of variables, we have obtained exact vortex solutions for this
case. We have investigated the NDE in two representations:  spherical and
spinor. In the spherical presentation we obtain optical vortices with spin
$l=1$ and in the spinor representation vortices with spin $j=1/2$. These
vortices are Lorenz invariant, and periodically pulsating with the group
velocity of solutions of the NME.


\begin{thebibliography}{100}

\bibitem{SW}
G. A. Swartzlander, Jr., C. T. Law, Phys. Rev. Lett {\bf 69},
2503-2506 (1992).

\bibitem{CH}
J. A. Christou "Optical vortices in beam propagation
dynamics" Australian National University, March (1999).

\bibitem{SIL}
Y. Silberberg, Opt. Lett.{\bf 15}, 1282-1284(1990).

\bibitem{ED}
D. E. Edmundson and R. H. Enns, Opt.Lett., {\bf 18}, 1609-1611
(1993).

\bibitem{MLD}
R. Mc. Leod, K. Wagner, and S. Blair, Phys. Rev. A {\bf 52}, 3254
(1995).

\bibitem{LW}
L. T. Law, G. A. Swartzlander, Jr., Chaos Solitons Fractals,
{\bf4},1759-1766(1994).

\bibitem{AH}
"Elecromagnetism and em waves', A. I. Achiezer
and I. A. Achiezer, Moskva, Vychaia Chkola, 1985.

\bibitem{LV}
 ''Nonlinear spin waves'', B. C. Lvov, Moskva, Nauka, 1987

\bibitem{DAC}
''Quantum mechanic'', A. Dacev, Nauka Izkustvo, 1973.

\bibitem{BR}   A. O. Barut,
''Geom. and Algebr. Aspects of Nonlinear Field Theory'', {\bf 37},(1989)

\end{thebibliography}
\end{document}